\documentclass[12pt,epsf]{article}
\usepackage{graphicx}
\usepackage{epsfig}
\usepackage{latexsym}
\begin{document}

\def\a{\alpha}
\def\b{\beta}
\def\c{\varepsilon}
\def\d{\delta}
\def\e{\epsilon}
\def\f{\phi}
\def\g{\gamma}
\def\h{\theta}
\def\k{\kappa}
\def\l{\lambda}
\def\m{\mu}
\def\n{\nu}
\def\p{\psi}
\def\q{\partial}
\def\r{\rho}
\def\s{\sigma}
\def\t{\tau}
\def\u{\upsilon}
\def\v{\varphi}
\def\w{\omega}
\def\x{\xi}
\def\y{\eta}
\def\z{\zeta}
\def\D{\Delta}
\def\G{\Gamma}
\def\H{\Theta}
\def\L{\Lambda}
\def\F{\Phi}
\def\P{\Psi}
\def\S{\Sigma}

\def\o{\over}
\def\beq{\begin{eqnarray}}
\def\eeq{\end{eqnarray}}
\def\lsim{\mathrel{\rlap{\lower4pt\hbox{\hskip1pt$\sim$}}
    \raise1pt\hbox{$<$}}}              
\def\gsim{\mathrel{\rlap{\lower4pt\hbox{\hskip1pt$\sim$}}
    \raise1pt\hbox{$>$}}}               
\newcommand{\vev}[1]{ \left\langle {#1} \right\rangle }
\newcommand{\bra}[1]{ \langle {#1} | }
\newcommand{\ket}[1]{ | {#1} \rangle }
\newcommand{\EV}{ {\rm eV} }
\newcommand{\KEV}{ {\rm keV} }
\newcommand{\MEV}{ {\rm MeV} }
\newcommand{\GEV}{ {\rm GeV} }
\newcommand{\TEV}{ {\rm TeV} }
\def\diag{\mathop{\rm diag}\nolimits}
\def\Spin{\mathop{\rm Spin}}
\def\SO{\mathop{\rm SO}}
\def\O{\mathop{\rm O}}
\def\SU{\mathop{\rm SU}}
\def\U{\mathop{\rm U}}
\def\Sp{\mathop{\rm Sp}}
\def\SL{\mathop{\rm SL}}
\def\tr{\mathop{\rm tr}}

\def\IJMP{Int.~J.~Mod.~Phys. }
\def\MPL{Mod.~Phys.~Lett. }
\def\NP{Nucl.~Phys. }
\def\PL{Phys.~Lett. }
\def\PR{Phys.~Rev. }
\def\PRL{Phys.~Rev.~Lett. }
\def\PTP{Prog.~Theor.~Phys. }
\def\ZP{Z.~Phys. }

\newcommand{\non}{\nonumber}
\newcommand{\bea}{\begin{eqnarray}}  \newcommand{\eea}{\end{eqnarray}}
\newcommand{\la}{\left\langle} \newcommand{\ra}{\right\rangle}
\def\lrf#1#2{ \left(\frac{#1}{#2}\right)}
\def\lrfp#1#2#3{ \left(\frac{#1}{#2}\right)^{#3}}


\baselineskip 0.7cm

\begin{titlepage}

\begin{flushright}
UT-08-25\\
IPMU 08-0053
\end{flushright}

\vskip 1.35cm
\begin{center}
{\large \bf
    Gauge Mediation with Sequestered Supersymmetry Breaking
}
\vskip 1.2cm
$^{(a)}$Satoshi Shirai, $^{(b)}$Fuminobu Takahashi, $^{(a,b)}$T. T. Yanagida and $^{(a)}$Kazuya Yonekura
\vskip 0.4cm

{\it  $^{(a)}$Department of Physics, University of Tokyo,\\
     Tokyo 113-0033, Japan}
\vskip 0.1in
{\em $^{(b)}$Institute for the Physics and Mathematics of the Universe,\\
University of Tokyo, Chiba 277-8568, Japan}

\vskip 1.5cm

\abstract{ Gauge mediation models have two drawbacks, that is, the
  so-called $\mu$-problem and a lack of predictability of the
  gravitino dark matter abundance. We show that conformal sequestering
  in the supersymmetry breaking sector offers attractive solutions to
  both problems. The correct mass scale of the $\mu$ and $B_\mu$ terms
  is generated by taking the gravitino mass of ${\cal O}(100)\,$GeV
  without causing the flavor-changing neutral-current  problem. 
  Moreover, a large anomalous dimension of the supersymmetry
  breaking field naturally realizes the small stau and neutralino mass
  difference required for the coannihilation to work yielding the
  right dark matter abundance.  }
\end{center}
\end{titlepage}

\setcounter{page}{2}

\section{Introduction}
Gauge-mediated supersymmetry (SUSY) breaking (GMSB) models~\cite{GM}
are very attractive, since those models can naturally solve the
flavor-changing neutral-current (FCNC)  problem in the SUSY
standard model (SSM).  This is because non-renormalizable operators at
the Planck scale $M_{\rm PL}$ are irrelevant for generating the soft
masses in gauge mediation. However, the GMSB models have two
drawbacks.  First, the origin of the so-called $\mu$ term is not clear
at all. If it is induced by the Planck suppressed operators, the $\mu$
parameter becomes of the order of the gravitino mass $m_{3/2}$, which,
however, is too small for the successful electroweak symmetry
breaking. The reason for this is that the gravitino mass is required
to satisfy $m_{3/2}<1$ GeV in order to suppress FCNC in the GMSB
scenario, provided that the non-renormalizable operators at the Planck
scale induce generic squark and slepton masses of order of the
gravitino mass.  Second, for such a light gravitino mass, it is the
gravitino that is a candidate for dark matter (DM) in the universe,
since the lightest SUSY particle in the SSM is not stable, and decays
into the gravitino.  The density of the gravitino depends crucially on
the reheating temperature after inflation, and hence we lose a
predictability of the DM density in the universe without knowledge of
the inflation dynamics.

We see that the above two problems originate from the small gravitino
mass. Thus, if we increase the gravitino mass up to ${\cal O}(100)$
GeV, both problems can be simultaneously solved.  In general, for the
gravitino mass of ${\cal O}(100)$ GeV, the soft masses for squarks and
sleptons given at the Planck scale induce too large FCNC. However,
this is not always the case. In this paper we show that, if the
conformal sequestering occurs in the SUSY breaking sector~\cite{CS,
  IINSY}, the above two problems are naturally solved in the GMSB
models without causing the FCNC problem.

In the present model we consider a parameter region where the lightest
neutralino is lighter than the gravitino and hence the stable lightest
SUSY particle (LSP).  Surprisingly enough, the present model naturally
predicts the small mass difference between the lightest neutralino and
the stau required for the coannihilation to work yielding the correct
DM density in the present universe~\cite{CoAN}. We would like to
stress that a large anomalous dimension of the SUSY breaking field $S$
is crucial to realize the coannihilation region naturally.

This paper is organized as follows. In Sec.~\ref{sec:2} we review a
model for the conformal sequestering, and see how the $\mu$-problem is
solved. In Sec.~\ref{sec:3} we study the neutralino DM density in
detail. We discuss the cosmological implications of our scenario in
Sec.~\ref{sec:4}. The last section is devoted for conclusions.

\section{A model for conformal sequestering of SUSY breaking}
\label{sec:2}
Let us consider the conformal sequestering of SUSY breaking, which
offers a natural solution to the $\mu$-problem as we will see at the
end of this section.  We first review a model of conformal
sequestering which was proposed in Ref.~\cite{IINSY}.  While we focus
on the model for concreteness, any models of conformal sequestering
containing a singlet SUSY breaking field $S$ may work as well.

\subsection{A hidden sector model}
We consider the IYIT SUSY breaking model~\cite{IYIT}, which is based
on an $SP(N)$ gauge theory with $2N+2$ chiral superfields $Q^i$
transforming in the fundamental representation of the gauge
group. Here, $i=1,\cdots,2N+2$ is the flavor index and we suppress the
gauge indices for simplicity. We also introduce
$\frac{1}{2}(2N+2)(2N+1)$ gauge singlet chiral fields,
$S_{ij}=-S_{ji}$. The tree level superpotential of this theory is
given by
\beq
W\;=\;h S_{ij} Q^i Q^j.
\eeq
Here we have assumed an $SU(2N+2)$ global symmetry~\footnote{In this
  paper we neglect subtlety regarding quantum gravitational effects on
  global symmetry.}  which acts on the indices $i,j$ in the
hidden sector, for simplicity.  This global symmetry is also imposed
on the K\"ahler potential as an exact symmetry for conformal
sequestering to work properly, because such operators that correspond
to conserved currents are not sequestered~\cite{CS}.  However, we can
relax this exact symmetry to an $SP(N+1)$, a subgroup of the
$SU(2N+2)$.  We will come back to this point later.

This theory exhibits a quantum deformation of the moduli
space~\cite{Seib}, and the low energy effective superpotential is
given by
\beq
W_{\rm eff}\;=\;X({\rm Pf}(Q^iQ^j) - (\L_{\rm SUSY})^{2N+2}) + h S_{ij}Q^iQ^j,
\eeq
where $X$ is a Lagrange multiplier and $\L_{\rm SUSY}$ is a dynamical
scale of the gauge theory, around which SUSY is broken.  The equation
of motion of $X$ requires ${\rm Pf}\vev{Q^iQ^j} = (\L_{\rm
  SUSY})^{2N+2}$. Then singlet fields $S_{ij}$ have $F$-term of order
$F_S \sim h \vev{QQ} \sim h (\L_{\rm SUSY})^2$, and SUSY is broken.

\begin{table}[t]
\begin{center}
\begin{tabular}{c|ccc} 
& $SP(N)$ & $SP(N')$ & $SP(N')$ \\ \hline
$Q \times 2(N+1)$ & $\Box_{2N}$ & ${\bf 1}$ & ${\bf 1}$ \\ \hline
$Q'_1$ &$\Box_{2N}$&$\Box_{2N'}$&${\bf 1}$ \\
$Q'_2$ &$\Box_{2N}$&${\bf 1}$&$\Box_{2N'}$ \\ \hline
$S_{ij}$ &${\bf 1}$&${\bf 1}$ &${\bf 1}$  
\end{tabular}
\caption{Matter contents of the model. $\Box_{2N}$ represents the fundamental representation of the gauge group $SP(N)$. This table is taken from
the Table 3 of Ref.~\cite{IINSY}. }
\label{tab1}
\end{center}
\end{table}

Let us introduce additional gauge symmetries and matter chiral
superfields so that the theory flows into a conformal fixed point
above the SUSY breaking scale.  We take an $SP(N) \times SP(N')^2
(=SP(N) \times SP(N')_1 \times SP(N')_2)$ model of Ref.~\cite{IINSY}
as a specific example.  In this model, there are matter chiral fields
$Q^i$ and $S_{ij} $ as above, and additional chiral fields $Q'_1$ and
$Q'_2$. The $Q'_{1(2)}$ transforms as a bi-fundamental representation
under $SP(N) \times SP(N')_{1(2)}$ and as a singlet under
$SP(N')_{2(1)}$.
See Table \ref{tab1}.  We take the superpotential of this model to be
\beq
W\;=\;h S_{ij} Q^i Q^j + m(Q'_1 Q'_1 + Q'_2 Q'_2),
\eeq
where $m$ is a mass parameter of $Q'$ at the Planck scale
$M_{PL} \simeq 2.4 \times 10^{18}~\GEV$. (The mass parameter $m$ is not equal to the physical mass of
$Q'$, $m_{\rm phys}$, because of a large anomalous dimension of $Q'$.)
If $N$ and $N'$ are appropriately chosen, we can expect (or can
explicitly show in the cases that we can use perturbation) that this
theory flows into a nontrivial fixed point~\cite{IINSY}.
  
Basic picture of this model is as follows.
As we lower a renormalization scale $\m_R$ from the Planck scale
$M_{PL}$, the theory enters conformal regime at some scale $M_*$,
which we assume to be much larger than $m_{\rm phys}$, but slightly
smaller than the Planck scale.  For $m_{\rm phys} \lsim \m_R \lsim
M_*$, the coupling constants of the theory are almost fixed at a
conformal fixed point, and the conformal sequestering occurs. For the
energy scale below the mass of $Q'$, i.e., $\m_R \lsim m_{\rm phys}$,
we can integrate out the massive fields $Q'$, and the theory becomes
identical to the IYIT model, and SUSY is broken at $\mu_R\simeq
\Lambda_{\rm SUSY}$ close to $m_{\rm phys}$.

Next let us discuss the suppression of higher dimensional operators in
a K\"ahler potential.  From a point of view of low energy effective
field theory, it is expected that there are higher dimensional terms
in a K\"ahler potential, which couple the hidden sector fields $A_i$($=Q$,
$Q'$ and $S$) and the visible sector fields $q_a$,
\beq
\D K = \frac{C_{ijab}}{M_{PL}^2} q^{\dagger}_a q_b A^{\dagger}_i A_j   \label{kahlerFCNC}
\eeq
with $C_{ijab}$ expected to be ${\cal O}(1)$. If $C_{ijab}$ is
generic, that is, if $C_{ijab}$ is not diagonal in the visible sector
flavor indices $a,b$, then these terms lead to the severe FCNC
problem. Conformal sequestering can solve this problem by suppressing
the terms in $\D K$ by renormalization group flow from the scale $M_*$
to the physical mass scale of $Q'$, $m_{\rm phys}$.
The suppression factor is roughly given by $(\m_R/M_*)^{\b'}$, where
$\b'={\q \b(\a)}/{\q \a}$ is a derivative of a beta function
$\b(\a)=\m_R (d\a / d \m_R)$ with respect to a coupling constant
$\a=g^2/4\pi$ of the theory~\footnote {Actually, the suppression
  factor is determined by the smallest eigenvalue of the matrix $(\q
  \b_k / \q \a_l)$ if there are more than one coupling constant.  In
  that case, $\b'$ of this section should be regarded as the smallest
  eigenvalue. See Ref.~\cite{IINSY} for details.}.  So, if $\b'$ is
sufficiently large, we expect a large suppression when we take the
energy scale $\m_R$ equal to the physical mass scale of $Q'$, $m_{\rm phys}$.

\begin{table}[Ht]
\begin{center}
\begin{tabular}{c|c|c}
&${\g_Q, \g_{Q'}, \g_S}$&$\b'$ \\ \hline
$SP(3) \times SP(1)^2$&{-1, -1, 2}& non-perturbative \\
$SP(5) \times SP(3)^2$&{-0.8, -0.8, 1.6}& non-perturbative \\
$SP(7) \times SP(5)^2$&{-0.7, -0.7, 1.4}& non-perturbative \\
$SP(13) \times SP(7)$&{-0.2, -0.8, 0.4}& 0.06 \\ 
$SP(20) \times SP(11)$&{-0.1, -0.8, 0.2}& 0.04 \\
\end{tabular}
\caption{Values of $\g$ and $\b'$. This table is taken from the Table 4 of Ref.~\cite{IINSY}. $\b'$ is the lowest eigenvalue of
the matrix $M$ of Ref.~\cite{IINSY}. }
\label{tab2}
\end{center}
\end{table}

The soft scalar masses of the visible fields receive contribution from
Eq.~(\ref{kahlerFCNC}),
\beq
\D m^2_{\rm vis} \sim C \left( \frac{m_{\rm phys}}{M_*} \right)^{\b'} m_{3/2}^2,
\eeq
where $m_{3/2}$ is the gravitino mass and $C$ collectively represents
$C_{ijab}$.  Since we will take $m_{3/2} = {\cal O}(100)~\GEV$ in our
scenario, $m_{\rm phys}$ is ${\cal O}(10^{10})\,$GeV. We also assume
$M_* \lsim M_{PL} \simeq 2.4 \times 10^{18}~\GEV$. The ratio of $\D
m_{\rm vis }$ to $m_{3/2}$ is then given by
\beq
\left( \frac{\D m_{\rm vis}}{m_{3/2}} \right)^2 \sim C \left(10^{-8} \cdot \frac{m_{\rm phys}}{10^{10}~\GEV} \cdot \frac{10^{18}~\GEV}{M_*} \right)^{\b'}.
\eeq
For $C = {\cal O}(1)$ and a relatively large value of $\b'$, the
ratio is small enough to satisfy the constraints from FCNC.
Note that phenomenological constraints from FCNC are rather mild
compared to the case of anomaly mediation ($m_{3/2} = {\cal
  O}(100)~\TEV$) due to the smaller gravitino mass.

A large anomalous dimension $\g_S$
of $S$, $\g_S \gsim 1$, will play a crucial role to account for the right
DM abundance as we see in the next section.  From Table \ref{tab2}, we see that we have $\g_S = 2$,
$\g_S = 1.6$ and $\g_S = 1.4$ for the cases of $SP(3) \times SP(1)^2$,
$SP(5) \times SP(2)^2$ and $SP(7) \times SP(3)^2$, respectively. In
those cases, we cannot calculate the precise values of $\b'$ because
the gauge and Yukawa couplings are very large and we cannot use
perturbation. Thus in this paper we simply assume that the suppression
is large enough to be consistent with FCNC constraints.

There are other higher dimensional operators in the K\"ahler
potential, which must be suppressed as well.  First, there are terms
linear in $S$, such as $S q^{\dagger} q/M_{PL}$.  These terms are
actually suppressed by a factor of $(m_{\rm phys}/M_*)^\frac{\g_S}{2}$
and therefore negligible. Second, there are terms which are cubic or
quartic in the hidden sector fields, e.g.,
\beq
\frac{1}{M_{PL}^4} (Q'Q') (Q'^{\dagger} Q'^{\dagger}) q^{\dagger} q.  \label{quartic_op}
\eeq
This term is suppressed by $1/M_{PL}^4$, and so, it may seem that this
is also negligible at a first glance. But in fact, terms like
Eq.~(\ref{quartic_op}) are dangerous. To see this, consider the case of
$SP(3) \times SP(1)^2$, in which the anomalous dimensions of $Q$, $Q'$
and $S$ are \beq \g_Q=-1,~~~~~\g_{Q'}=-1,~~~~~\g_{S}=2, \eeq and those
operators $QQ$ and $Q'Q'$ saturate the unitarity bound of conformal
field theory~\cite{Mack}.  If there is no vertex renormalization, the
anomalous dimension of $(Q'Q')(Q'^{\dagger}Q'^{\dagger})$ is
$\g_{Q'Q'Q'^{\dagger}Q'^{\dagger}}/2=-2$, and the operator
(\ref{quartic_op}) is enhanced by a factor of $(M_*/m_{\rm phys})^2$.
If $M_* \sim M_{PL}$, the operator is effectively suppressed only by
$1/M_{PL}^2$, and is not negligible.  Indeed, if $Q'Q'$ has
non-vanishing $F$-term $F_{(Q'Q')} \neq 0$ which is comparable with
$F_S$, the operator (\ref{quartic_op}) leads to too large flavor
dependent soft masses of the visible sector, causing a FCNC
problem. Actually, however, it is suppressed by a factor
$(M_*/M_{PL})^2$ if $M_*\lsim M_{PL}$, and so, the FCNC problem can be
avoided if we take $M_* \ll M_{PL}$~\footnote{
T.~T.~Y. thanks Y.~Nakayama and M.~Ibe for serious discussions on this
problem.
}.  In the numerical analysis of the next section, we take $SP(3)
\times SP(1)^2$ model with $M_* \sim 10^{16}~\GEV$ as an example.  In
that case $(M_*/M_{PL})^2 \sim 10^{-4}$, and there is no FCNC problem.
We will neglect the soft masses induced from Eq.~(\ref{quartic_op}) in the
following analysis. 

\subsection{Coupling the hidden sector to messenger fields}
In our GMSB model, we introduce a Yukawa interaction between a singlet
field and messenger fields in the superpotential,
\beq
W\;=\;\l S \P {\bar \P} \label{mess-int},
\eeq
where $\P$ and ${\bar \P}$ are the messenger superfields charged under
standard model gauge groups.  We would like to make two comments
concerning the introduction of this term.

First, we need to single out one singlet field $S$ from the singlets
$S_{ij}$. This can be done by reducing the global symmetry of the
theory from $SU(2N+2)$ to $SP(N+1)$.  Then, $S_{ij}$ can be decomposed
as $S_{ij}=S'_{ij} + S R_{ij}$, where $R_{ij}$ is the $SP(N+1)$
invariant tensor and $(R^{-1})^{ij}S'_{ij}=0$. We then have to allow
two different couplings $h_1$ and $h_2$ in the superpotential, \beq
W\;=\;h_1 SR_{ij}Q^iQ^j + h_2 S'_{ij}Q^iQ^j.  \eeq It is reasonable to
assume that also in this case the theory flows into a conformal fixed
point which is stable in the infrared.  At the fixed point, vanishing
of $\b$ functions of $h_1$ and $h_2$ requires
$\g_S+2\g_Q=\g_{S'_{ij}}+2\g_Q=0$, and we have $\g_S=\g_{S'_{ij}}$.
This suggests that the fixed point of this theory is the same as in
the case that we impose $SU(2N+2)$ symmetry.  In other words, there
is an enhanced $SU(2N+2)$ symmetry at the fixed point~\footnote{This
  is an example of the ``emergent symmetries'' discussed in
  Ref.~\cite{SS}}.

Conserved currents $A^{\dagger} T^\a A$ ($T^\a$ are generators of
$SU(2N+2)$ and $A=\{S_{ij},~Q^i\}$) have vanishing anomalous
dimensions, and so, operators like $\D K = C_{ab\a} q_a^\dagger q_b
A^{\dagger} T^\a A $ are not sequestered.  Then we have to worry about
non-sequestering of such conserved currents~\cite{CS,SS}.  The
$SU(2N+2)$ adjoint representation can be decomposed into symmetric and
traceless-anti-symmetric representation of $SP(N+1)$ (trace is taken
by contracting indices with $R_{ij}$), and there is no trivial
representation.  If we impose the $SP(N+1)$ symmetry on the K\"ahler
potential, therefore, there is no conserved current which can appear
in the K\"ahler potential.  Thus non-sequestering of conserved
currents does not occur in our case.

In fact, it may even be possible that we impose no symmetry on the
K\"ahler potential at all.  The K\"ahler potential of $A$ with
dangerous conserved current operators is
\beq
K\;=\;A^\dagger A + \e_\a A^\dagger T^\a A,
\eeq
where $\e_\a=C_{ab\a}q_a^\dagger q_b$. For the purpose of calculating
the soft masses, we can suppose that $q_a$ are constants.  Then we can
transform the hidden fields $A$ as
\beq
A \rightarrow \left(1-\frac{1}{2} \e_\a T^\a \right)A
\eeq
so that the K\"ahler potential becomes
\beq
K \rightarrow A^\dagger A + {\cal O}(\e^2).
\eeq
The point is that because this transformation corresponds to the
symmetry transformation of the whole theory, which is respected even
by the conformal symmetry breaking mass term of $Q'$, we can
completely transform away the visible fields
$\e_\a=C_{ab\a}q_a^\dagger q_b$ and no soft mass is generated.  For
more discussions on conserved currents, see Ref.~\cite{SS}.  The above
argument suggests that even if we do not impose any symmetry at all on
the K\"ahler potential, we may achieve conformal sequestering without
the danger caused by conserved currents.  The only requirement is that
the theory should flow into the infrared stable fixed point for
arbitrary Yukawa couplings $h^{ij}_{kl}S_{ij}Q^kQ^l$.

Second, there is a danger that introducing the coupling
(\ref{mess-int}) may significantly deform the original theory.  We
argue that this interaction is in fact harmless for the hidden sector
dynamics.  Suppose that the value of $\l$ in Eq.~(\ref{mess-int}) and
the standard model gauge couplings are not so large at the scale
$M_*$.  Then, the anomalous dimension of $\P$ and ${\bar
  \P}$, $\g_{\P}$, is small.  In this case the renormalization group equation of
$\l$ is given by
\beq
\m \frac{d}{d\m} | \l | = \left( \frac{\g_S}{2}+ \g_{\P} \right) | \l | \simeq \frac{\g_S}{2} | \l |  > 0.
\eeq
As we lower the energy scale $\m$, $\l$
becomes smaller and smaller, and so does the contribution of
Eq.~(\ref{mess-int}) to $\g_{\P}$.  The effect of the interaction
(\ref{mess-int}) to the hidden sector dynamics therefore becomes
totally negligible.  In other words, the operator of
Eq.~(\ref{mess-int}) is an irrelevant operator of renormalization
group flow.  Even if $\l$ is somewhat large at the scale $M_*$, at
least in the leading order of perturbation theory, the Yukawa coupling
gives positive contribution to $\g_{\P}$, and so, the relation
$\g_S+2\g_\P>0$ still holds. This fact makes the above discussion more
robust.

While $\l$ at the scale ${M_*}$ is naturally expected to be ${\cal
  O}(1)$, it gets suppressed at the SUSY breaking scale (and therefore at
  the messenger mass scale) due to strong
conformal dynamics.  The value of $\l$ at the scale $m_{\rm phys}$ is
given by
\beq
\l|_{m_{\rm phys}}\; \simeq\; \left( 10^{-8}\cdot \frac{m_{\rm phys}}{10^{10}~\GEV}\cdot 
\frac{10^{18}~\GEV}{M_*} \right)^{\frac{\g_S}{2}} \l_0,
\label{lsup}
\eeq
where we have defined the value of $\l$ at the scale $M_*$ as $\l_0
\equiv \l|_{M_*}$.
As we will see in the next section, the smallness of $\l|_{m_{\rm
    phys}}$ is essential for the coannihilation to occur in a wide
parameter region of $B_\m/\m$.

\subsection{The origin of $\mu$ and $B_\mu$ terms}
\label{sec:2-3}
Before closing this section, let us explain the origin of $\m$ and
$B_\m$ terms in our model.  Although in our model the soft masses of
the visible sector are generated by gauge mediation, the $\m$ term and
$B_\m$ term are generated by supergravity effects~\cite{Inoue:1991rk}. 
We assume that there is some global $U(1)_R$ symmetry in the theory, under which
Higgs doublets are neutral.  Then, arbitrary $\m$ term is forbidden by
this symmetry, but the following terms in the K\"ahler
potential~\footnote{The K\"ahler potential $K$ in this section is that
  in the conformal frame of supergravity. The usual K\"ahler potential in
  the Einstein frame $K_{\rm sugra}$ is related to this K\"ahler
  potential by $K_{\rm sugra} = -3 M_{PL} ^2 \log(1- K/3
  M_{PL}^2)$. Conformal sequestering occurs in the conformal frame of
  supergravity.}  and the superpotential are allowed:
\beq
K &\supset& c H_u H_d  + {\rm h.c.},\\ 
W &\supset& c' (m_{3/2})^* H_u H_d. \label{super_mu}
\eeq
The interaction (\ref{super_mu}) is allowed because $(m_{3/2})^*
\propto W_0$ has $U(1)_R$ charge 2, where $W_0$ is a constant term in
the superpotential~\footnote{The phase of the gravitino mass $m_{3/2}$
  is determined as follows.  In the compensator formalism of
  supergravity, the Lagrangian of the compensator field
  $\F=1+F_{\F}\h^2$ is ${\cal L} = \int d\h^2 d{\bar \h}^2 [-3M_{PL}^2
    \F^{\dagger} \F \exp(-K_{\rm sugra}/3M_{PL}^2)]+ \int d\h^2 W \F^3
  +{\rm h.c.}  = -3M^2_{PL}|F_{\F}|^2+3W_0 F_{\F}+{\rm h.c.}+\cdots$
  where dots denote terms irrelevant for the vev of $F_\F$ and the
  lowest component of $\F$ is gauge fixed to be $1$.  By solving the
  equation of motion of $\F$, we have $\vev{F_{\F}}=W^*_0/M_{PL}^2$.
  We define the phase of $m_{3/2}$ such that $m_{3/2} \equiv
  \vev{F_{\F}} = W_0^* / M_{PL}^2$.}.  We expect that $c$ and $c'$ are
${\cal O}(1)$ parameters. In the compensator formalism of
supergravity~\cite{compensator}, we have to put the compensator field,
$\F$, so the K\"ahler potential and the superpotential become
\beq
K &\supset& c H_u H_d \frac{\F^{\dagger}}{\F} + {\rm h.c.}, \\ 
W &\supset& c' (m_{3/2})^* H_u H_d \F.
\eeq  
Substituting a vacuum expectation value (vev) $\vev{\F} = 1 + m_{3/2}
\h^2$ and integrating over $d\h^2$ and/or $d{\bar \h}^2$, we obtain
the $\m$ and $B_\m$ terms
\beq
\m &=& (c+c')(m_{3/2})^* \label{eq:mu},\\
B_\m &=& (-c+c')|m_{3/2}|^2. \label{eq:b}
\eeq 
The correct mass scale of $\mu$ and $B_\mu$ can be generated for the
gravitino mass of ${\cal O}(100)\,$GeV with $c, \,c' = {\cal O}(1)$.
Note that the FCNC  problem is absent thanks to the conformal
sequestering of the SUSY breaking.  Other terms such as A-terms which
are generated by the anomaly mediation are suppressed by one-loop factor.  
An alternative solution to the $\mu/B_\mu$
problem was proposed in Refs.~\cite{Roy:2007nz}.

\section{A sequestered GMSB model and the neutralino relic density}
\label{sec:3}

We consider a simple GMSB model, where a SUSY breaking field $S$
couples to $N_5$ pairs of messenger chiral superfields, $\P$ and
$\bar{\P}$, which transform as ${\bf 5}$ and ${\bf 5}^*$ under the
$SU(5)_{\rm GUT}$:
\beq
W \;=\; \lambda S\Psi{\bar \Psi}+M \Psi{\bar \Psi},
\eeq
where $M$ is the messenger mass and $\l$ is set to be the value at
$m_{\rm phys}$ throughout this section, i.e., $\lambda = \l|_{m_{\rm phys}}$.  
A priori $\l$ is a free parameter, however, in our
scenario, $\l$ is naturally very small: $\lambda \simeq
10^{-6}-10^{-7}$ (see Eq.~(\ref{lsup})).  The SUSY breaking field $S$
develop a vev $\langle S \rangle = \theta^2 F_S$, which is related to
the gravitino mass as $|F_S| = \sqrt{3} m_{3/2} M_{PL}$, assuming that
the SUSY breaking is dominated by $F_S$.

In the GMSB models, the SSM gaugino masses are generated from loop
diagrams of the messengers. At the one-loop level, gaugino masses are
given by
\begin{equation}
M_{a} \;=\; \frac{N_5\alpha_a}{4\pi}\Lambda_{eff} g(x),
\label{eq:gaugino_mass}
\end{equation}
where we have defined $\Lambda_{eff} =\l F_S/M$, $x=\l F_S/M^2$, 
 and 
\begin{equation}
g(x)\; =\; \frac{1}{x^2}[(1+x)\log(1+x)+(1-x)\log(1-x)].
\end{equation}
Here $a = 1,2, 3$ labels $U(1), SU(2)$ and $SU(3)$ in the SSM, respectively, 
and we use the normalization $\alpha_1=5 \alpha _{\rm EM}/(3 \cos^2\theta_{W})$.
The soft scalar masses arise at the  two loop level, and given by
\begin{equation}
m^2_{\phi_i}\;=\;2N_5\Lambda_{eff} ^2 \sum_a \left(\frac{\alpha_a}{4\pi}\right)^2 C_a (i) f(x), \label{eq:scalar_mass}
\end{equation}
where $C_a(i)$ are Casimir invariants for the visible particles $\phi_i$ 
($C_1(i) = 3Y_i^2/5$)
and
\begin{equation}
f(x) \;=\; \frac{1+x}{x^2}\left[ \log(1+x) - 2 {\rm Li}_2(x/[1+x]) + \frac{1}{2} {\rm Li}_2(2x / [1+x]) \right] + (x \rightarrow -x).
\end{equation}
For $x<1$, both $f(x)$ and $g(x)$ are ${\cal O}(1)$.  
We see that $m_{\f_i} \simeq M_a = {\cal O}(1)~\TEV$ is realized for $\L_{eff}={\cal O}(10^5)~\GEV$.

Since the above
expressions for the soft masses are given at the messenger scale, one
should solve the visible sector renormalization group (RG) equation to
get the on-shell masses and mixing matrices.  To this end, we have
used the program {\verb SOFTSUSY } 2.0.18~\cite{Allanach:2001kg},
setting ${\rm sgn}(\mu) = +1$.  In our analysis, we choose
$B_{\mu}/\mu$ at the messenger scale as a free parameter and
$\tan\beta$ (ratio of two higgs expectation values) is determined by
given parameters.  This is because we can naturally expect
$B_{\mu}/\mu = {\cal O}(m_{3/2})$ at the SUSY breaking scale from
Eqs.~(\ref{eq:mu}) and (\ref{eq:b}). Notice that $(B_{\mu}/\mu)_{\rm
  SUSY~breaking}=(B_{\mu}/\mu)_{\rm messenger}$ at least at the
one-loop level of RG equations, assuming that the other SSM soft parameters vanish above the messenger scale.

\begin{figure}[t!]
\begin{center}
\epsfig{file=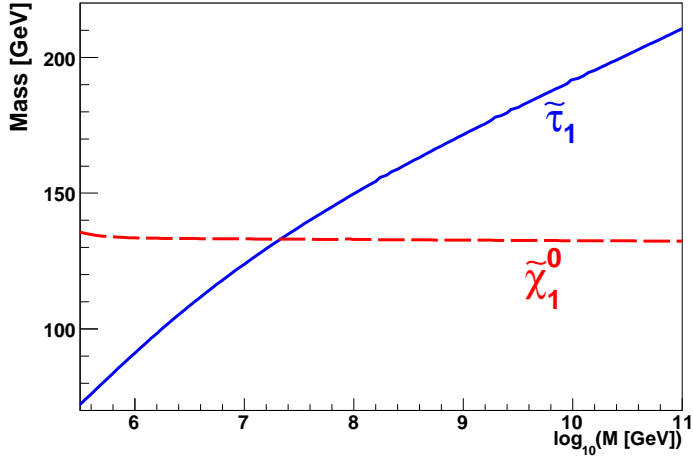,clip,scale=.5}
\caption[]{$m_{\tilde{\tau}_1}$ and $m_{\tilde{\chi}^0_1}$'s dependence on $M$ 
for $\Lambda_{eff} = 10^4$ GeV, $N_5=1$ and
$(B_{\mu}/\mu)_{\rm messenger}=0$ GeV.
 }
\label{fig:mass1}
\end{center}
\end{figure}

In the present GMSB model, the lighter stau ($\tilde{\tau}_1$),  the
lightest neutralino ($\tilde{\chi}^0_1$) or the gravitino becomes LSP.  
We mainly consider a parameter region where the neutralino, $\tilde{\chi}^0_1$,
is the LSP and hence a candidate of the DM.
Which of the two particles, $\tilde{\tau}_1$ or $\tilde{\chi}^0_1$, becomes the LSP mainly depends on the number of the
messenger ($N_5$), the mass of the messenger $(M)$, and $B_{\mu}/\mu$.
As can be seen in Eqs.~(\ref{eq:gaugino_mass}) and
(\ref{eq:scalar_mass}), the gauginos become heavier as increasing $N_5$.  
The stau becomes heavier for a larger mass of the messenger, while the
gaugino masses are almost independent of the messenger mass.  Hence,
in the case of the heavy messenger, the lighter stau mass $m_{\tilde{\tau}_1}$ 
tends to become heavier than the mass of the lightest neutralino,  $m_{\tilde{\chi}^0_1}$ 
(see. Fig.~\ref{fig:mass1}).

Larger $\tan\beta$ implies stronger tau's Yukawa coupling.
Therefore, the stau becomes lighter through left-right mixing and RG effects for the larger $\tan\beta$.
The value of $(B_{\mu}/\mu)_{\rm messenger}$ is connected to the value of $\tan\beta$.
In general, a smaller $B_{\mu}/\mu$ leads to a larger $\tan\beta$.

\subsection*{Coannihilation}
Let us first calculate a naively expected range for the messenger scale in our scenario.
The value of $\l$ is related to the value of $\l_0 \equiv \l|_{M_*}$ by
\beq
\l \sim \left(\frac{\sqrt{F_S}}{M_*}\right)^{\g_S/2} \l_0,
\eeq
where we have substituted $\sqrt{F_S} \sim m_{\rm phys}$, assuming the Yukawa coupling $h$
is of order unity. Using the relations $\L_{eff} = \l F_S/M$ and $m_{3/2} = F_S/\sqrt{3}M_{PL}$, we obtain~\footnote{
From the SSM soft parameters, we can rotate away all but one complex phases in our model. This remaining complex phase has a potential danger for
the SUSY CP problem. An accurate bound for this phase from the CP constraint depends on details of the spectrum
of the SUSY parameters. In fact, we see the CP problem becomes milder in the region where $B_\mu/\mu$ is smaller than the wino mass.
Here we neglect the remaining phase and take all parameters to be real in this paper, for simplicity.}
\beq
M \sim 10^{15-3\g_S}~\GEV \times \l_0 \left( \frac{\L_{eff}}{10^5~\GEV} \right)^{-1} \left( \frac{m_{3/2}}{10^2~\GEV} \right)^{1+\g_S/4} 
\left(\frac{10^{16}~\GEV}{M_*}\right)^{\g_S/2}.
\label{eq:expmass}
\eeq
If we adopt  the hidden sector model of $SP(3) \times SP(1)^2$ and $M_* \sim 10^{16}~\GEV$, Eq.~(\ref{eq:expmass}) leads to
\beq
M \sim 10^9~\GEV \times \l_0 \left( \frac{\L_{eff}}{10^5~\GEV} \right)^{-1} \left( \frac{m_{3/2}}{10^2~\GEV} \right)^{3/2}.
\label{mess_mass}
\eeq
Hence,  the messenger mass is expected to be ${\cal O}(10^9)~\GEV$ in the model,
unless the value of $\l_0$ is fine-tuned to be much smaller than unity.

\begin{figure}[t!] 
\begin{center}
\epsfig{file=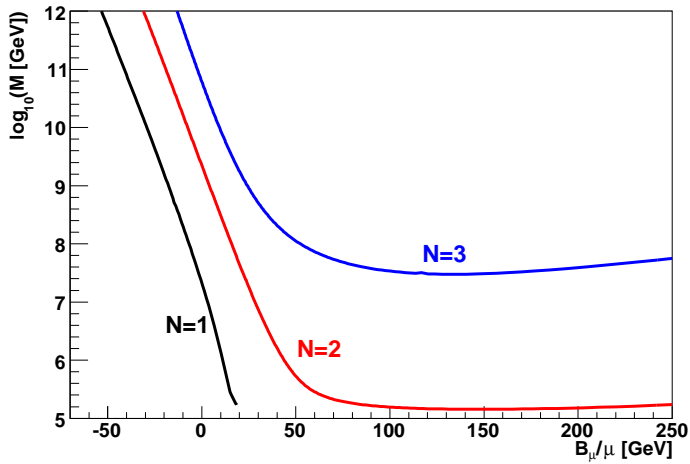,clip,scale=.5}
\caption[]{$(B_{\mu}/\mu)_{\rm messenger}$ dependence of the messenger mass 
which realizes $m_{\tilde{\tau}_1} = m_{\tilde{\chi}^0_1}$.
We set $\Lambda_{eff} = 10^5, 5\times10^4, 3\times10^4$ GeV for $N_5=1, 2, 3$, respectively.
}
\label{fig:mass2}
\end{center}
\end{figure}

Next, let us discuss the parameter region in which the coannihilation takes place.
Roughly speaking, the coannihilation occurs when the lighter stau mass becomes very close to
the lightest neutralino mass~\cite{CoAN}.
In Fig.~\ref{fig:mass2} we show the relation between $M$ and $B_\m/\m$ when $m_{\tilde{\tau}_1} = m_{\tilde{\chi}^0_1}$ is met.
From the figure, we can see that required coannihilation occurs for a wide region of $B_\m/\m$ if 
the messenger mass $M$ is approximately $10^8~\GEV$ for $N_5=3$. 
Such value of $M$ is realized naturally in our model for $\l_0 = {\cal O}(10^{-1})$ (see  Eq.~(\ref{mess_mass})).
Note also that we can obtain the value $N_5=3$ not only by introducing 3 pairs of messengers which 
transform as ${\bf 5}$ and ${\bf 5}^*$, but also by introducing  one pair of messengers transforming as ${\bf 10}$ and ${\bf 10}^*$. 
In the case that $N_5=2$, $M = {\cal O}(10^5)~\GEV$ is required for the coannihilation to occur with a
wide parameter region of $B_\m/\m$. This is achieved by a rather small value of $\l_0 = {\cal O}(10^{-4})$.

\subsection*{Relic Density}
\begin{figure}[h!]
\begin{tabular}{cc}
\begin{minipage}{0.5\hsize}
\begin{center}
\epsfig{file=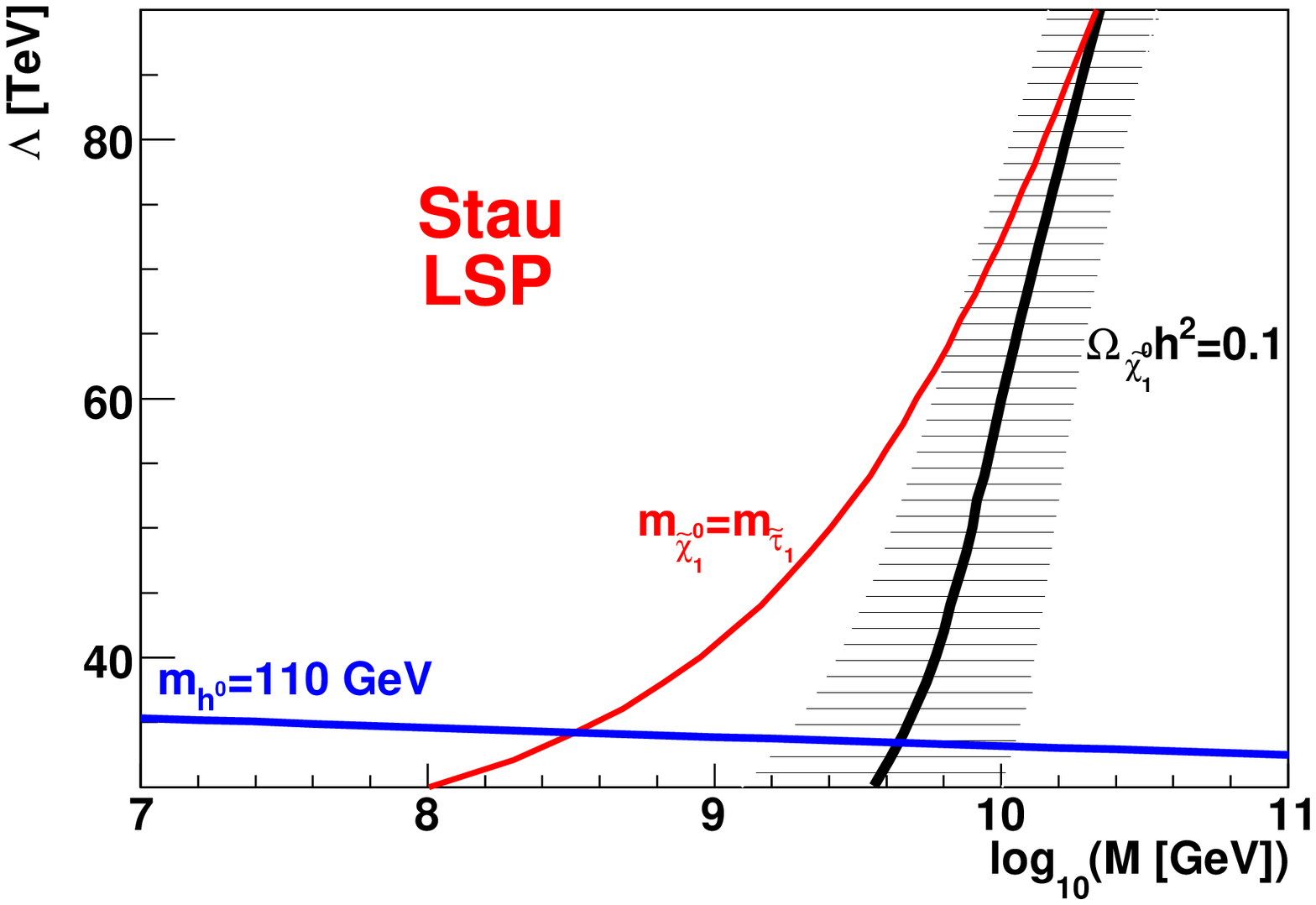,clip,scale=.45}
(a)
\end{center}
\end{minipage}

\begin{minipage}{0.5\hsize}
\begin{center}
\epsfig{file=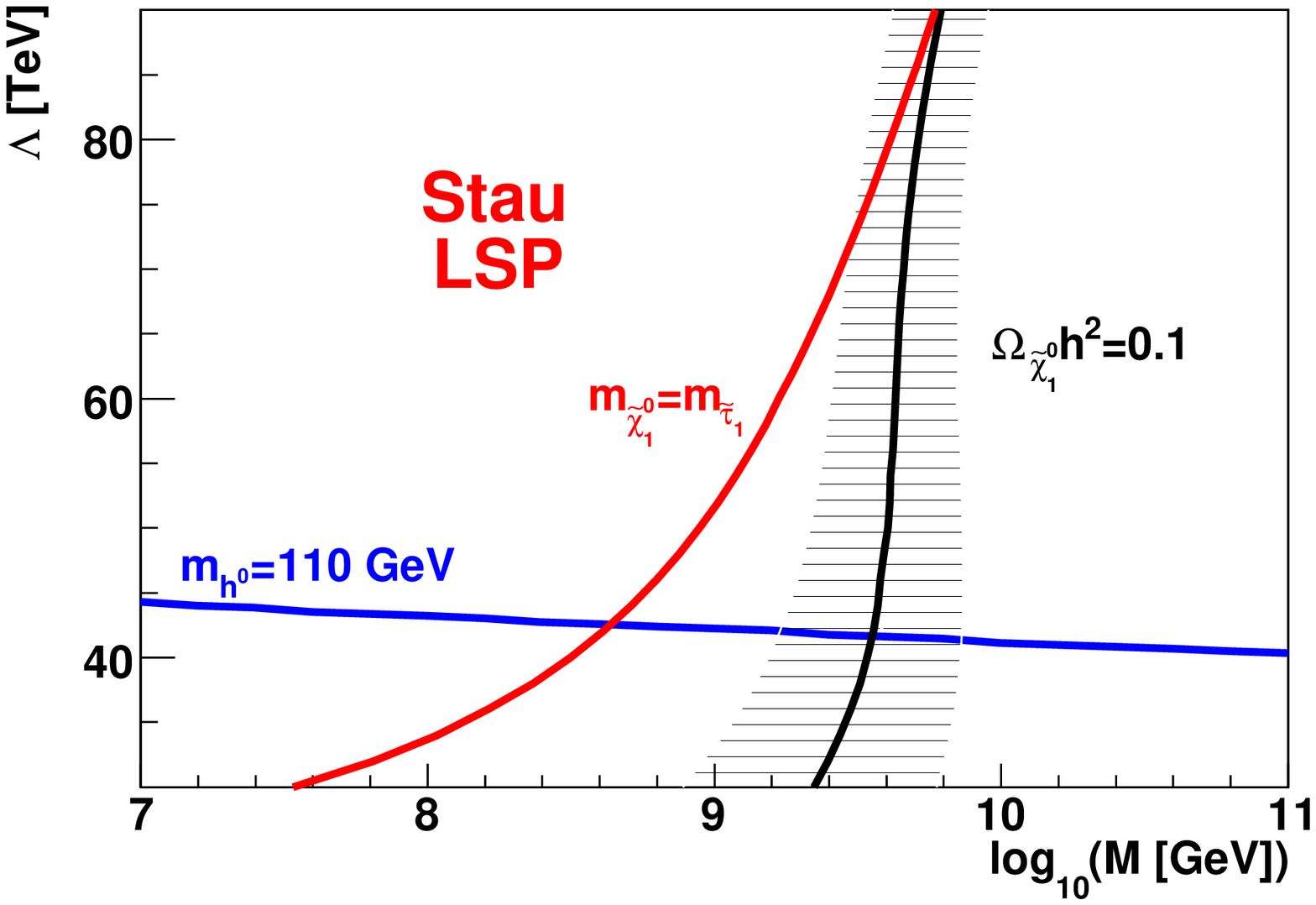,clip,scale=.45}
(b)
\end{center}
\end{minipage}
\end{tabular}
\caption[]{Contour plot of $\Omega_{\tilde{\chi}^0_1} h^2$ on the $(M, \Lambda_{eff})$ 
plane for $N_5=3$ and (a) $B_{\mu}/\mu= 50$ GeV, (b) $B_{\mu}/\mu= 100$
GeV at the messenger scale.
The black line represents $\Omega_{\tilde{\chi}^0_1} h^2=0.1$, and the shaded region shows the
ambiguity from the AMSB effect.
Here, we set $m_{3/2}=500$ GeV.
The red line represents $m_{\tilde{\chi}^0_1}=m_{\tilde{\tau}_1}$. 
On the left  side of this red line, the stau becomes the LSP.
The blue line represents $m_{h^0}=110$ GeV.
Above this blue line, $m_{h^0}$ becomes larger than $110$ GeV.
}
\label{fig:N3}
\end{figure}

From the viewpoint of naturalness, the case that $N_5=3$ seems to be
most interesting, since it naturally predicts a suitable value of
messenger mass (Eq. (\ref{mess_mass})) for $B_{\mu}/\mu={\cal O}(m_{3/2})$ 
(see Fig.~\ref{fig:mass2}).  Now we show that this
model actually predicts the correct abundance of the neutralino DM. 
 In Fig.~\ref{fig:N3}, a contour plot of $\Omega_{\tilde{\chi}^0_1} h^2$ on the
$(M, \Lambda_{eff})$ plane is shown.  Here we set $(B_{\mu}/\mu)_{\rm
  messenger}= 50~\GEV,~100~\GEV$ for Fig. \ref{fig:N3}-(a) and (b),
respectively.  We have used the program {\verb MicroOmegas }
2.2~\cite{micromegas} to estimate the cold dark matter density.  
Here, we set $m_{3/2}=500$ GeV, and
the red and blue lines represent $m_{\tilde{\chi}^0_1}=m_{\tilde{\tau}_1}$ and
 $m_{h^0}=110$ GeV, respectively.
We can see that $\Omega_{\tilde{\chi}^0_1} h^2 \simeq 0.1$ is realized for
$M=10^{9}-10^{10}$ GeV.  This value of the messenger mass is nothing
but the expected one from Eq. (\ref{mess_mass}). 
In Fig.~\ref{fig:N3} we have taken into account the anomaly-mediation (AMSB)
effects~\cite{AMSB} to the SUSY breaking soft masses for the SSM particles.

\section{Cosmology}
\label{sec:4}
Let us discuss the cosmological implications of our scenario. In the previous section,
we have seen that the neutralino LSP, instead of the gravitino, can naturally 
account for the observed DM abundance. It does
not necessarily mean, however, that the cosmological abundance of the gravitino
is totally negligible. In fact, gravitinos can be produced thermally directly from the hot plasma,
and non-thermally from the inflaton decay. It is known that the gravitinos 
can induce a severe cosmological problem~\cite{Weinberg:zq,Krauss:1983ik,BBNwX_OLD}.

The  abundance of the gravitinos produced from thermal scatterings 
is given by~\cite{Bolz:2000fu,Kawasaki:2004yh,Pradler:2006qh}
\begin{eqnarray}
    \label{eq:Yx-new}
    Y_{3/2}^{(TH)} &\simeq& 
    1.9 \times 10^{-12}\left[ 1+ 
    \left(\frac{m_{\tilde{g}_3}^2}{3m_{3/2}^2}\right)\right]
    \left( \frac{T_{ R}}{10^{10}\ {\rm GeV}} \right)
    \nonumber \\ 
    & \times & 
    \left[ 1 
        + 0.045 \ln \left( \frac{T_{ R}}{10^{10}\ {\rm GeV}} 
        \right) \right]
    \left[ 1 
        - 0.028 \ln \left( \frac{T_{R}}{10^{10}\ {\rm GeV}} 
        \right) \right],
\end{eqnarray}
where $T_R$ is the reheating temperature and $m_{\tilde{g}_3}$ is the gluino
running mass evaluated at the reheating. 
Moreover, the gravitinos are generically produced by
the inflaton decay, if the inflaton has a non-vanishing vev (more precisely, a non-vanishing
linear term in the K\"ahler potential) at the potential minimum~\cite{
Kawasaki:2006gs,Endo:2006qk,Endo:2007sz,Endo:2007ih}. For an inflaton mass lighter than the
SUSY breaking scale, the gravitino pair production becomes efficient~\cite{Kawasaki:2006gs} (see also Refs.~\cite{Endo:2006zj,Dine:2006ii}). 
On the other hand, 
for the inflaton mass heavier than the SUSY breaking scale, the gravitinos are produced from
the inflaton decay into the hidden gauge sector~\cite{Endo:2007ih, Endo:2007sz}.
 The abundance of the non-thermally produced gravitinos is given by
\beq
 Y_{3/2}^{(NT)} 	&\simeq & 7 \times 10^{-11}\, x \lrfp{g_*}{200}{-\frac{1}{2}} \lrfp{\la \phi \ra}{10^{15}{\rm GeV}}{2}
 	\lrfp{m_\phi}{10^{12}{\rm GeV}}{2} \lrfp{T_R}{10^6{\rm GeV}}{-1},
\label{NT}	
\eeq
where $g_*$ counts the relativistic degrees of freedom, $\la \phi \ra$
is the inflaton vev, and $m_\phi$ the inflaton mass. Here $x$ is a
numerical coefficient given by
\beq
x &=& \left\{
\begin{array}{cc}
1 & {\rm ~~~for~~~}m_\phi < \Lambda_{\rm SUSY} \\
~10^{-3} {\rm~~\sim~~}10^{-1}& {\rm ~~~for~~~}m_\phi > \Lambda_{\rm SUSY} 
\end{array}
\right.,
\label{eq:valuex}
\eeq
The precise value of $x$ depends on the detailed structure of the SUSY
breaking sector.

In our scenario, the gravitino mass is set to be of the order of
$100\,$GeV to generate the $\mu$-term of a right magnitude, and the
gravitino is not the LSP and therefore unstable.  For such unstable
gravitino, the total gravitino abundance must satisfy
\beq
Y_{3/2} \;\equiv\;  Y_{3/2}^{(TH)} +  Y_{3/2}^{(NT)}
\;\lsim\;{\cal O}(10^{-16}),
\label{bbn}
\eeq
in order not to spoil the success of the big bang nucleosynthesis
(BBN)~\cite{Kawasaki:2004yh,Kohri:2005wn,Kawasaki:2008qe,Jedamzik:2004er}.
Substituting Eq.~(\ref{eq:Yx-new}) into Eq.~(\ref{bbn}), we obtain an upper
bound on $T_R$:
\beq
T_R &\lsim& {\cal O}(10^6) {\rm \,GeV}.
\label{uptrth}
\eeq
It is non-trivial for an inflation model to satisfy the bound on
$T_R$~\cite{Endo:2006qk,Endo:2007sz}.  Indeed, it rules out the smooth
hybrid inflation~\cite{Lazarides:1995vr} as well as a part of the
parameter space of the hybrid inflation~\cite{Copeland:1994vg}.  In
addition, the non-thermal gravitino production excludes most of the
inflation models such as the new~\cite{Izawa:1996dv,Asaka:1999jb} and
hybrid inflation.  Note that one cannot avoid the gravitino
overproduction simply by reducing the reheating temperature due to the
peculiar dependence of $Y_{3/2}^{(NT)}$ on $T_R$.

Among possible solutions to the (non-thermal) gravitino
overproduction, the simplest one is to suppress the inflaton vev by
imposing a symmetry on the inflaton. As a concrete example, let us
consider a chaotic inflation model with a $Z_2$
symmetry~\cite{Kawasaki:2000yn}.  In this model, we assume that the
K\"ahler potential $K(\phi,\phi^\dag)$ is invariant under the shift of
$\phi$,
\begin{equation}
  \phi \rightarrow \phi + i\,A,
  \label{eq:shift}
\end{equation}
where $A$ is a dimension-one real parameter.  We also impose a $Z_2$
symmetry: $\phi \rightarrow - \phi$.  Then, the K\"ahler potential is
given by
\beq
K(\phi+\phi^\dag)=  \frac{1}{2} (\phi+\phi^\dag)^2 + \cdots,
\eeq
 where we have dropped a linear term of $(\phi + \phi^\dag)$ which is
 forbidden by the $Z_2$ symmetry.  We introduce a small breaking term
 of the shift symmetry in the superpotential to generate a potential
 for the inflaton:
\begin{equation}
  W(\phi,\psi) = m_{\rm inf} \,\phi \,\psi, 
  \label{eq:mass}
\end{equation}
where we have introduced a new chiral multiplet $\psi$ charged under
the $Z_2$ symmetry: $\psi \rightarrow -\psi$. The inflaton mass $m_{\rm inf}
\simeq 2\times10^{13}$\,GeV represents the breaking scale of the shift
symmetry, and reproduces the density fluctuations of the right
magnitude.  The imaginary part of $\phi$ is identified with the
inflaton field $\varphi \equiv \sqrt{2} {\rm \,Im}[\phi]$, and the
scalar potential is given by
\beq
V(\varphi,\psi) \;\simeq\; \frac{1}{2}m_{\rm inf}^2 \varphi^2 + m_{\rm inf}^2 |\psi|^2,
\eeq
after the real part of $\phi$ settles down to the minimum.  For
$\varphi \gg M_{PL}$ and $|\psi| < M_{PL}$, the $\varphi$ field
dominates the potential and the chaotic inflation takes place (for
details see Ref.~\cite{Kawasaki:2000yn}).  Since the linear term in
the K\"aher potential is absent thanks to the $Z_2$ symmetry, the
non-thermal gravitino production does not occur.

In order to induce the reheating into the visible sector, we consider
the following interactions:
\beq
W_{\rm int} \;=\; \frac{k}{2}\, \phi N N + \frac{1}{2} M_N NN,
\label{int-N}
\eeq
where $N$ is a right-handed neutrino chiral multiplet.  The $Z_2$
symmetry is explicitly broken by those interactions, and we will later
discuss how small the breaking should be. For $m_{\rm inf} \gg 2 M_N$, the decay
rate is given by
\beq
\Gamma_N \;\simeq\; \frac{k^2}{32\pi}  m_{\rm inf}.
\eeq
Assuming that the reheating occurs mainly through the decay into the
right-handed (s)neutrinos, the reheating temperature is given by
\beq
\label{tr2}
T_R \;\simeq\; 2 \times 10^6{\rm \,GeV} \lrf{k}{10^{-8}} \lrfp{m_{\rm inf}}{2\times 10^{13} {\rm GeV}}{\frac{1}{2}},
\eeq
where we have defined the reheating temperature as
\beq
\label{eq:def-Tr}
T_R \;\equiv\; \lrfp{\pi^2 g_*}{10}{-\frac{1}{4}} \sqrt{\Gamma_N M_{PL}}.
\eeq
The non-thermal leptogenesis occurs in this
case~\cite{Fukugita:1986hr,Asaka:1999yd, Lazarides:1993sn}, and the
resultant baryon asymmetry is given by
\beq
\frac{n_B}{s} \;\simeq\; 1 \times 10^{-10}\,\lrf{k}{10^{-8}} \lrf{M_N}{10^{13}{\rm GeV}} \lrfp{m_{\rm inf}}{2\times 10^{13} {\rm GeV}}{-\frac{1}{2}}
 \lrf{m_{\nu_3}}{0.05{\rm eV}} \delta_{\rm eff},
\eeq
where $m_{\nu_3}$ is the heaviest neutrino mass and $ \delta_{\rm eff}
\leq 1$ represents the effective $CP$-violating phase.  Note that a
right amount of the baryon asymmetry is generated for $k \sim 10^{-8}$
and $M_N \sim 10^{13}$\,GeV, corresponding to $T_R \sim 10^{6}$GeV being
marginally compatible with the constraint (see Eq.~(\ref{uptrth})).

Now let us discuss the $Z_2$ symmetry breaking.  We may interpret the
first term in Eq.~(\ref{int-N}) breaks both the shift and $Z_2$
symmetries, with an assumption that $NN$ is even under the $Z_2$
symmetry.
Then, we may attribute the smallness of $k \sim 10^{-8}$ to the
breaking of the $Z_2$ symmetry, while the inflaton mass $m_{\rm inf}/M_{PL} \sim
10^{-5}$ represents the typical magnitude of the shift symmetry
breaking~\footnote{ Alternatively, we can interpret that the first
  term in Eq.~(\ref{int-N}) breaks the shift symmetry while the second
  term breaks the $Z_2$ symmetry, by assigning a $Z_2$ odd charge to
  $NN$.}.  Since the $Z_2$ symmetry is explicitly broken, a linear
term in the K\"ahler potential is induced at one-loop level: $\delta K
\; \sim\; 1/(16 \pi^2) k M_N^* \phi + {\rm h.c.}$.  Or, since the $Z_2$
symmetry is not a true symmetry of the theory, we may expect the
presence of a linear term, $K = \tilde{c} M_{PL}\, (\phi + \phi^\dag)$ with $\tilde{c}
\sim 10^{-8}$, from the beginning.  Our concern is if such a tiny
$Z_2$ breaking leads to the gravitino overproduction again.  To
satisfy the BBN constraint (\ref{bbn}), the coefficient $c$ must be
suppressed as
\beq
\tilde{c} &\lsim &{\cal O}(10^{-7}) \cdot x^{-1/2},
\eeq
where we have substituted $\la \phi \ra \simeq \tilde{c} M_{\rm PL}/\sqrt{2}$
and $m_\phi = m_{\rm inf}$ into Eq.~(\ref{NT}). Therefore, for $\tilde{c} \sim k \sim
10^{-8}$, we can avoid the non-thermal gravitino overproduction
problem~\footnote{
If the gravitino is the LSP, a right amount of the gravitino DM
can be produced for $k \sim \tilde{c} \sim m_{\rm inf} \sim 10^{-5}$ (in the Planck unit),
and the thermal leptogenesis becomes possible. The BBN bound
can be avoided by including tiny violation of the $R$-parity~\cite{Takayama:1999pc,Buchmuller:2007ui},
and the decay of the unstable gravitino may explain the
anomalies observed by HEAT and EGRET~\cite{Ibarra:2007wg,Ishiwata:2008cu}.
}.  In addition, the non-thermal leptogenesis is also possible.

Lastly let us make a comment on the Polonyi problem~\cite{Polonyi}.
If the SUSY breaking field $S$ has a non-vanishing linear term in the
K\"ahler potential, the initial position of $S$ during inflation is
generically deviated from the origin~\cite{Ibe:2006am}.  Such a linear
term may not exist at tree level, but it is necessarily generated due
to the coupling to the messenger fields (\ref{mess-int}) at one-loop
level.  If the deviation were large, the SUSY breaking field might
produce too many gravitinos. Fortunately, due to the large anomalous
dimension of the SUSY breaking field, $S$, the messenger mass scale
is suppressed, and so does the linear term. Therefore
there is no Polonyi problem in our scenario~\cite{Endo:2007cu}

\section{Conclusions}
\label{sec:5}
In this paper we have pointed out that a conformal sequestering of the SUSY breaking
can naturally solve the two problem inherent in the gauge mediation; the $\mu/B_\mu$
problem and the lack of predictability of the gravitino DM abundance. 
First, since the dangerous higher dimensional operators in the K\"ahler potential 
are suppressed due to the conformal sequestering, we can increase the gravitino mass
up to ${\cal O}(100)\,$GeV without causing the FCNC problem. 
The correct mass scale of the $\mu$ and $B_\mu$ terms can be generated for 
such gravitino mass. Second, a large anomalous dimension of the SUSY breaking field
makes the messenger scale very small, which results in a small mass difference
between the neutralino and the stau, making the coannihilation to naturally occur.
We have also discussed the cosmological implications of our scenario. The unstable 
gravitino of a mass of $100$\,GeV suffers from a severe gravitino overproduction problem, but
we can find an example in which the problem is avoided and the right amount of the
baryon asymmetry is generated through the non-thermal leptogenesis.

\vspace{5mm}

{\it Acknowledgments:} The work of
S.S. is supported in part by JSPS Research Fellowships for Young
Scientists.  The work of T.T.Y. is supported in part by the
Grant-in-Aid for Science Research, Japan Society for the Promotion of
Science, Japan (No.\ 1940270). 
This work was supported in part by World Premier International
Research Center Initiative WPI Initiative), MEXT, Japan.

\end{document}